# Flow-field Analysis and Performance Assessment of Rotating Detonation Engines under Different Number of Discrete Inlet Nozzles

Sebastian Valencia[1,2], Andres Mendiburu[3], Luis Bravo[4], Prashant Khare[5,6], Cesar Celis[1,2]

[1]Mechanical Engineering Section, Pontificia Universidad Católica del Perú, Lima, 15088, Peru
[2]Flows, Particles, Combustion and Environment (FPCE), Pontificia Universidad Católica del Perú, Lima, 15088, Peru
[3]Universidade Federal do Rio Grande do Sul, Porto Alegre, RS 90040-060, Brazil
[4]US Army Research Laboratory, Aberdeen Proving Ground, MD 21005, US
[5]University of Cincinnati, Cincinnati, OH 45221-0070, US
[6]Hypersonics Laboratory, Digital Futures, University of Cincinnati, Cincinnati, OH 45221-0070, US

## Abstract

This study explores in depth rotating detonation engines (RDEs) fueled by premixed stoichiometric hydrogen/air mixtures through two-dimensional numerical simulations including a detailed chemical kinetic mechanism. To model the spatial reactant non-uniformities observed in practical RDE combustors, the referred simulations incorporate different numbers of discrete inlet nozzles. The primary focus here is to analyze the influence of reactant non-uniformities on detonation combustion dynamics in RDEs. By systematically varying the number of reactant injection nozzles (from 15 to 240), while maintaining a constant total injection area, the study delves into how this variation influences the behavior of rotating detonation waves (RDWs) and the associated overall flow field structure. The numerical results obtained here reveal significant effects of the number of inlets employed on both RDE stability (self-sustaining detonation wave) and performance. RDE configurations with a lower number of inlets exhibit a detonation front with chaotic behavior (pressure oscillations) due to an increased amount of unburned gas ahead of the detonation wave. This chaotic behavior can lead to the flame extinguishing or decreasing in intensity, ultimately diminishing the engine's overall performance. Conversely, RDE configurations with a higher number of inlets feature smoother detonation propagations without chaotic transients, leading to more stable and reliable performance metrics. This study uses high-fidelity numerical techniques such as adaptive mesh refinement (AMR) and the PeleC compressible reacting flow solver. This comprehensive approach enables a thorough evaluation of critical RDE characteristics including detonation velocity, fuel mass flow rate, impulse, thrust, and reverse pressure waves under varying reactant injection conditions. The insights derived from the numerical simulations carried out here enhance the understanding of the fundamental processes governing the performance of RDE concepts.

# 1    Introduction

Combustion processes involve a sequence of highly exothermic chemical reactions set in motion by the interplay of fuel, oxidizer, and an ignition source (Zhou et al., 2016). This intricate process gives rise to a combustion wave that radiates outwards, engaging with reactants to release stored energy from chemical bonds, thereby converting it into thermal and kinetic energy within the resulting products (Lee, 2008). This released energy sustains the propagation of combustion waves, which in turn drives various engine processes. Combustion can be classified into deflagration and detonation, with detonation processes exhibiting accelerated reactions and superior thermal efficiency, which can exceed conventional deflagrative engine cycles by up to 19% (Anand & Gutmark, 2019).

The pursuit of detonation-based engines has led to significant advancements in this field, resulting in the development of standing detonation engines (SDEs), pulse detonation engines (PDEs), and rotating detonation engines (RDEs). In particular, RDEs have garnered attention due to their streamlined structure, self-sustaining detonation wave propagation, high frequency, and remarkable specific power outputs (Wolański, 2013). In RDEs typically featuring an annular combustion chamber, fuel and oxidizer are introduced through orifices, either in pre-mixed or non-premixed configurations (J. Shaw et al., 2021). The initiation of the detonation wave involves a deflagration-to-detonation transition (DDT) preceding entry into the chamber, where fuel is consumed, leading to thrust generation (J. Shaw et al., 2021).

Recent years have witnessed an increased interest in rotating detonation engines, with extensive research progress documented in detailed reviews (Anand & Gutmark, 2019; J. Shaw et al., 2021; Raman et al., 2023; Xie et al., 2020). These review works summarize the latest advancements and research findings in this rapidly evolving field. As highlighted by Liu et al. (2020), one common observation across several studies relates to the banded distribution of reactants in front of detonation waves, a phenomenon attributed to the discrete injection mode utilized in experiments, such as holes or slits. Consequently, when studying both stability and performance of RDEs, the numerical modeling of reactant discrete injection has become a useful approach.

The way an injector design handles mixing is crucial in determining the strength and structure of detonations, so this has been a major area of research interest within the scientific community under non premixed and premixed configurations (Fujii et al., 2017; Mikoshiba et al., 2024). In the premixed context, Wang et al. (Wang et al., 2020) conducted a study on kerosene/air RDEs, investigating different injection area ratios (1, 0.8, and 0.4), representing the relationship between the inlet area and the total area of the inlet plus the wall, for a single nozzle, thereby maintaining a fixed number of inlet nozzles. Their findings revealed that at low area ratios, burned gas tends to remain trapped in the triangular fresh mixture layer, leading to a reduction in detonation velocity. Similarly, Liu et al. (Liu et al., 2020) studied different inlet area ratios (0.4, 0.6, 0.72, and 1) in RDEs accounting for H2/air mixtures. They observed that when the ratio falls below 1.0, the reactants in front of the detonation waves exhibit a discrete banded distribution, causing reverse-compression waves in the flow-field. Additionally, they noticed an increase in both the specific impulse and the specific thrust as the inlet area ratio increased. In contrast, Fujii et al. (2017) focused on varying the number of inlet nozzles under premixed and non-premixed RDEs configurations and C2H4/O2 mixtures. Their investigation showed that increasing the nozzle interval results in more burned gas accumulating in front of the detonation wave, with minimal changes in propagation velocity. However, their study primarily focused on detonation velocity, leaving a gap in the understanding of the impact of varying the number of inlet nozzles on detonation wave stability and other RDE performance parameters. Further research work is thus essential to gain a comprehensive understanding of these issues.

To address the questions indicated above, accounting for two-dimensional RDEs and stoichiometric H2/air mixtures, in this work a comprehensive series of numerical simulations is carried out. These numerical simulations involve a varying number of discrete inlet nozzles (15, 30, 60, 120, and 240), while keeping a fixed ratio of 3/2 between the total inlet area and the wall area. In particular, the reactive Euler equations are solved here using an adaptive mesh refinement (AMR) based solver specifically developed for modelling compressible reacting flows, PeleC (Henry de Frahan et al., 2022). In addition, the detailed chemical kinetic mechanism developed by Li et al. (2004), which features 9 chemical species and 21 chemical reactions, is used to model chemical non-equilibrium processes. The main goal of this study is to investigate the influence of the number of discrete inlet nozzles on the propagation velocity and self-sustained stability of the detonation wave, as well as on the overall performance of RDEs. Accordingly, the remainder of the work is structured as follows. Section 2 describes the mathematical and numerical models employed here, Section 3 discusses the main numerical results obtained in this work, and Section 4 summarizes some of the conclusions drawn from them.

## 2    Mathematical and Numerical models

In this section, the main features of both the mathematical and numerical models employed in this work are highlighted.

### 2.1 Governing equations

All numerical simulations carried out in this work involve solving the unsteady, two-dimensional (2D) Euler equations coupled with chemical source terms accounting for chemical reactions. The choice to use the reactive Euler equations here was guided by previous works, including those by Chen et al. (2023), Fujii et al. (2017), and Wang et al. (2020), which also studied RDEs featuring discrete inlets using these equations. Neglecting viscous effects in the reactive Euler equations simplifies the numerical simulations by omitting detailed treatments of small-scale turbulent mixing and viscous dissipation. While these viscous effects are important for enhancing mixing at small scales and achieving a more homogenized reactant-product mixture before detonation wave propagation, their omission was a deliberate choice in this work. This was done to focus on capturing the main features of detonation wave dynamics, including shock interactions and chemical reactions effects, without the added complexity of a full Navier-Stokes formulation. Although the use of a Euler equations-based approach may lead to an underestimation of fine-scale mixing between reactants and product gases, it ensures that the core physical mechanisms responsible for wave propagation and instabilities are properly described. Therefore, this work exclusively relies on the Euler equations, omitting the incorporation of viscosity and thermal wall-effects into the numerical simulations. Accordingly, the transport equations solved here related to, respectively, mass, momentum, species, and energy – ignoring viscosity, thermal conduction, and mass diffusion, read as follows,

$$\frac{\partial}{\partial t}(\rho) + \nabla \cdot (\rho \vec{u}) = 0 \qquad (1)$$

$$\frac{\partial}{\partial t}(\rho \vec{u}) + \nabla \cdot (\rho \vec{u} \otimes \vec{u}) + \nabla p = 0 \tag{2}$$

$$\frac{\partial}{\partial t}(\rho Y_k) + \nabla \cdot (\rho \vec{u} Y_k) = \rho \dot{\omega}_k \tag{3}$$

$$\frac{\partial}{\partial t}(\rho E) + \nabla \cdot (\rho \vec{u} E + p \vec{u}) = \sum_k h_k \dot{\omega}_k \tag{4}$$

where $\rho, \vec{u}, Y_k, t, \dot{\omega}_k, h_k, p$, and $E$ stand for density, velocity, $k$-th chemical species mass fraction, time, k-th chemical species reaction rate, species formation enthalpy, pressure, and total energy, respectively. The system of equations solved is closed using the ideal gas equation of state, complemented by the 9 species, 21 reactions hydrogen-air detailed chemical kinetic mechanism developed by Li et al. (J. Li et al., 2004). Notice that the kinetic mechanism chosen has been developed for reactions involving pure hydrogen and is suitable for a wide range of operational conditions, including those at elevated pressures characterizing RDEs. Furthermore, this chemical kinetic mechanism was originally included into the PeleC software.

## 2.2 Solver and Numerical schemes

In this study, PeleC (Henry de Frahan et al., 2022), an AMR finite-volume solver for compressible reacting and non-reacting fluid flow simulations with complex geometry, was employed. The flow governing equations were closed using the ideal gas equation of state (EoS) from the PelePhysics submodule (Bell et al., 2022). PeleC uses for time-stepping an iterative scheme base on a spectral deferred correction approach (SDC). To address the discretization of hyperbolic fluxes, the second order unsplit piecewise parabolic hybrid PPM/WENO method with the 5th order WENO-Z scheme (Colella & Sekora, 2008) was employed. This hybrid strategy presents far better results in terms of capture of turbulent spectra rather than other PPM methods (Motheau & Wakefield, 2021). In addition, the integration of the stiff-chemistry ordinary differential equations (ODE) was performed using the third-party library SUNDIALS (Hindmarsh et al., 2005), that is, using CVODE-based integration (Cohen Y And & Hindmarsh, n.d.). Finally, the CFL (Courant–Friedrichs–Lewy) was fixed at 0.2 for all numerical simulations performed here.

## 2.3 Geometric configuration

Given the relatively small width of annular RDE chambers relative to their diameter, several previous studies (Chen et al., 2023; Fujii et al., 2017; Luan et al., 2022; Zhao et al., 2020) have employed two-dimensional computational domains instead of three-dimensional ones. Notice that the decision to employ a 2D domain in this work was driven by the need to reduce computational costs and to be able to perform a parametric study of fuel injector configurations within a manageable timeframe. Although this approach enables the exploration of fundamental effects associated with injector design, it is acknowledged that a comprehensive 3D analysis would offer a more detailed and accurate depiction of the associated flow dynamics, including wave curvature, three-dimensional mixing, and non-axisymmetric effects. In addition, this geometric simplification also neglects other crucial three-dimensional effects, such as wall boundary layers, lateral relief, vortex stretching, and channel curvature (Luan et al., 2022). Nevertheless, it is worth noticing that two-dimensional configurations can effectively replicate the main characteristics of the flow field structure and the propagation process of rotating detonation waves (Chen et al., 2023).

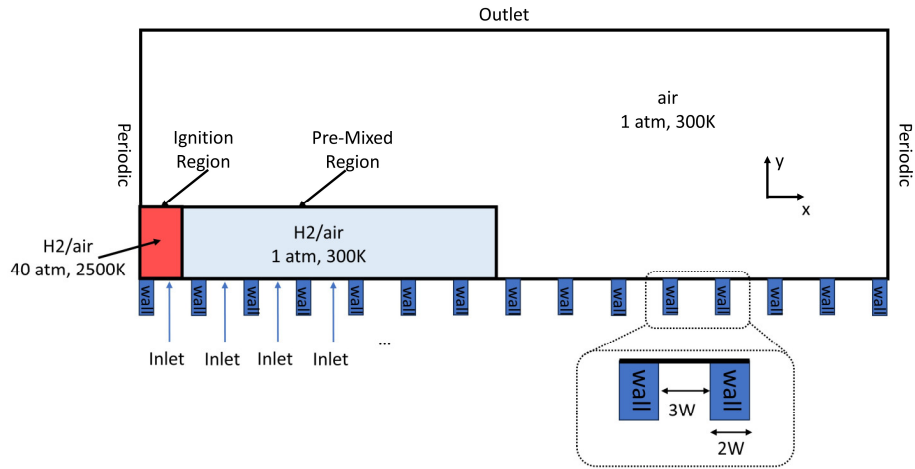

Figure 1. Two-dimensional computational domain and boundary conditions.

Consequently, similar to the domain studied by Chen et al. (2023), a two-dimensional rectangular channel was accounted for in this work as the computational domain. As illustrated in Figure 1, in terms of physical dimensions, this rectangular channel features 150 mm along the x-direction and 50 mm in the y-direction. The initial flow field is defined by saturating the computational domain with air at 300 K and 1 atm. Subsequently, a premixed region featuring 80 mm width and 12 mm height, and a stoichiometric H2/air mixture at the same temperature and pressure conditions, is also defined. Finally, to establish an initial propagating detonation front, a localized ignition region, initialized with 2500 K and 40 atm, which spans 1 mm width and 12 mm height, is employed.

## 2.4 Boundary conditions

The left and right boundaries of the two-dimensional domain are designated as periodic (Figure 1). This type of boundary condition ensures the dynamic evolution of the system through the continuous rotation of the detonation wave. At the outlet (top boundary of the computational domain, Figure 1), the boundary condition employed here depends on the flow Mach number. When the outlet flow is supersonic, the flow parameters at the ghost cells are equal to those characterizing the interior ones. However, when the outlet flow is subsonic, a specific treatment is applied, where the flow density and the three velocity components in the ghost cells are determined by extrapolating the values at the interior ones, whereas pressure is extrapolated in the ghost cells and set on the interior ones (Blazek J., 2015). To reduce shock reflections, similar treatments have been employed in previous works (Wang et al., 2020), as they ensure both a smooth transition at the outlet boundary and a reduced number of numerical stability issues in simulations involving subsonic outlet flows.

The boundary conditions at the inlet of the system in turn are characterized by an alternating arrangement of inlet and wall components. In the wall indeed, adiabatic no-slip boundary conditions are imposed. Additionally, the area ratio between the inlet and the wall is fixed at 3/2, while the total number of inlets varies from 15, 30, 60, 120, to 240. In accordance with the theory of gas isentropic expansion (Zhdan et al., 1990), the flow inlet conditions, including pressure ($P$), temperature ($T$), and velocity normal to the inlet ($v$), are determined based on the relationship between the total inlet pressure of the nozzle ($P_0$), the total temperature of the incoming flow ($T_0$), and the pressure extracted from the first cell in the interior field near the inlet ($P_w$). As described below, the referred relationship establishes three specific inlet conditions based on isentropic expansion through the micro-nozzles (Schwer & Kailasanath, 2011).

(i) No Injection. When $P_w \geq P_0$, the reactants cannot be injected into the chamber. The injection flow properties are then determined as follows.

$$P = P_w, \qquad T = T_0 \left(\frac{P}{P_0}\right)^{\frac{\gamma-1}{\gamma}}, \qquad v = 0 \tag{5}$$

(ii) Subsonic Injection. In the case that $P_0 > P_w > P_{cr}$, where $P_{cr} = P_0 \left(\frac{2}{\gamma+1}\right)^{\frac{\gamma-1}{\gamma}}$ is the critical pressure, the inlet flow is not choked and the injection flow properties are computed from,

$$P = P_w, \qquad T = T_0 \left(\frac{P}{P_0}\right)^{\frac{\gamma-1}{\gamma}},$$

$$v = \sqrt{\frac{2\gamma}{\gamma-1} R T_0 \left[1 - \left(\frac{P}{P_0}\right)^{\frac{\gamma-1}{\gamma}}\right]} \tag{6}$$

(iii) Sonic Injection. When $P_w \leq P_{cr}$, the flow conditions are computed from the critical flow ones, critical pressure $P_{cr}$ and critical temperature $T_{cr} = T_0 \left(\frac{2}{\gamma+1}\right)$. More specifically, as the flows at the inlets are choked, the injection flow properties are determined from,

$$P = P_{cr}, \qquad T = T_{cr}, \qquad v = \sqrt{\frac{2\gamma}{\gamma-1} R T_0} \tag{7}$$

Finally, similar to previous studies (Chen et al., 2023), the total inlet pressure $P_0$ and the total temperature $T_0$ are set in this work as being equal to 0.5 MPa and 300 K, respectively.

## 2.5 Computational grid

The computational domain employed in the numerical simulations carried out here was initially discretized using 600 points along the x-direction and 200 points along the y-direction, resulting in a total number of 120,000 square cells of 0.25 mm × 0.25 mm each. During the numerical simulations, the base grid was dynamically refined on-the-fly with two levels of AMR based on density and pressure gradients. Additionally, as illustrated in Figure 2, the entire flow injection region was refined with the same number of AMR levels, achieving a mesh resolution of 62.5 μm × 62.5 μm (0.0625 mm × 0.0625 mm) in specific areas of the domain.

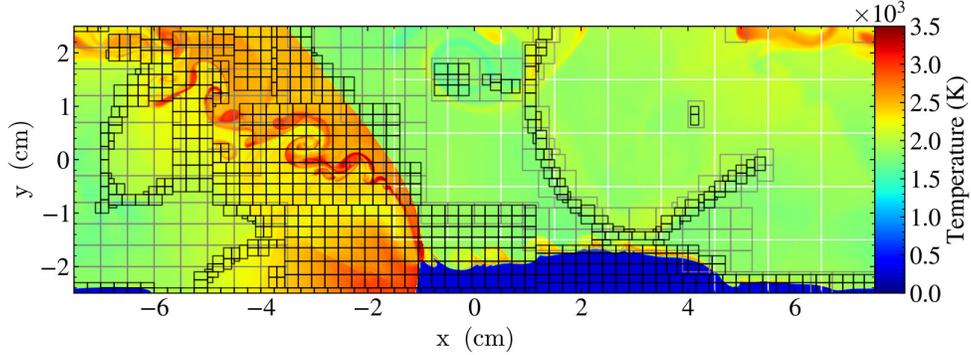

Figure 2. Computational mesh details including AMR for a continuous inlet configuration.

Figure 2 depicts indeed an instantaneous temperature field of a simulation conducted with a continuous injection configuration (where the total bottom area serves as inlet), highlighting an instant when the rotating detonation wave (RDW) is steady and self-sustaining. The black and gray boxes shown in this figure represent the regions where AMR was applied. Specifically, the black boxes indicate areas with the highest level of refinement, whereas the gray boxes indicate regions with the lowest level of refinement. This mesh resolution is expected to be sufficient to accurately capturing the flow characteristics in the camber. Notice that previous works have used similar mesh resolutions. For instance, Chen et al. (Chen et al., 2023) used mesh cell sizes of 0.1 mm, while Zhao et al. (Zhao et al., 2020) found that 0.2 mm allows properly predicting the main structures needed to perform RDE instability analyses. For the sake of clarity, the discussions about the quality of the meshes employed in this work are postponed to Section 3.1.

## 3    Results

This section summarizes the key findings of this study. Initially, an assessment of mesh quality is discussed, followed by an examination of the flow field structure in each reactive flow configuration studied here. Subsequently, a RDW stability analysis is presented for one of the flow configurations accounted for, along with the impact of reverse shock waves on the stable propagation of RDWs. Additionally, detonation velocity and engine performance analyses are also conducted to provide further insights into the associated physical phenomena.

## 3.1 Mesh sensitivity analysis

A mesh quality assessment was initially conducted using four different computational meshes, all of them featuring different levels of AMR. These meshes include (i) a mesh without AMR (AMR 0) ($\Delta_{min}$ = 0.25 mm), (ii) a mesh with 1 level of AMR (AMR 1) ($\Delta_{min}$ = 0.125 $mm$), (iii) a mesh with 2 levels of AMR (AMR 2) ($\Delta_{min}$ = 0.0625 mm), and (iv) a mesh with 3 levels of AMR (AMR 3) ($\Delta_{min}$ = 0.03125 mm). The purpose of this assessment was to evaluate the mesh resolution requirements for the flow configurations studied here. Accordingly, Figure 3 displays, for a continuous inlet configuration, the temperature fields computed using the four different meshes analyzed here. Notice that these AMR levels related results were obtained by starting with a stable numerical solution without any AMR level (AMR 0). Then, for each different AMR level case, the numerical simulations were conducted for 5 wave cycles, and the results shown for the different mesh resolutions accounted for correspond to the same time instant. Accordingly, the temperature contours shown in Figure 3 reveal that the four meshes analyzed in this work capture relatively well the basic flow structures, including the detonation wave, oblique shock wave, slip line, combustible premixture filling region, and the deflagration surface. In the case without AMR, the detonation wave shifts to the left, indicating a lower wave velocity. In contrast, the other cases show similar detonation wave positions, but the finer grids capture more effectively the characteristics of the vortical structures induced by flow instabilities along both the slip line and the deflagration surface.

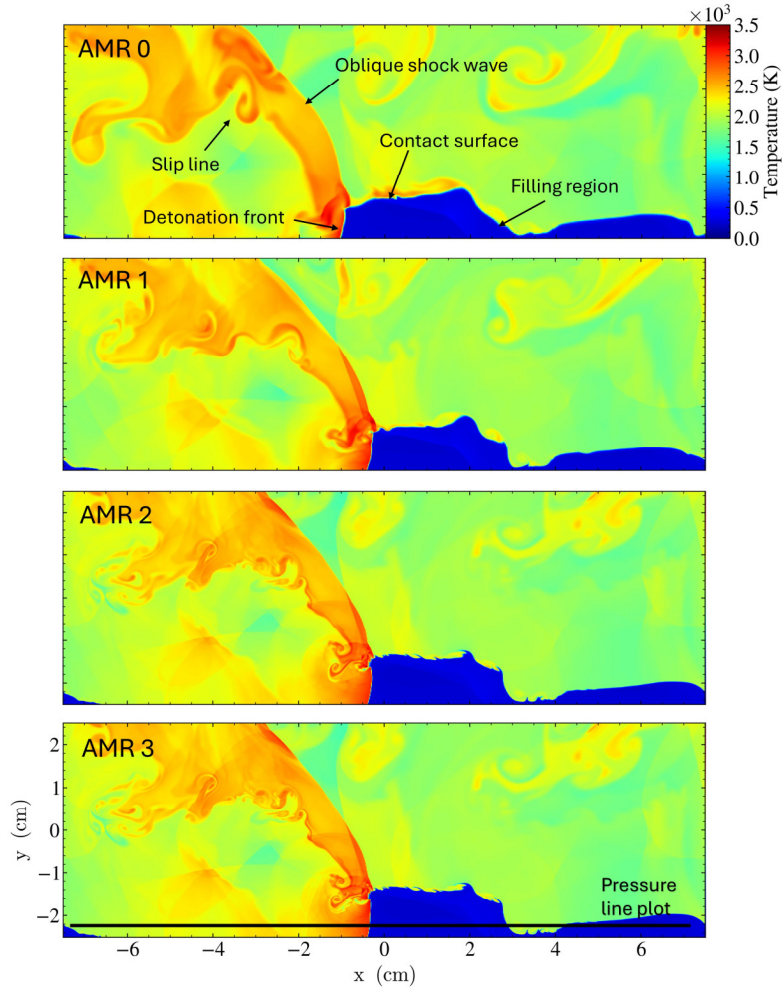

Figure 3. Temperature fields computed with four different meshes (AMR 0, AMR 1, AMR 2, and AMR 3) and a continuous inlet configuration.

To complement these qualitative results, as illustrated in Figure 4, pressure data was extracted along the x-direction at y = 0.5 mm (for details about the x and y directions see Figure 2). The pressure profile obtained from the referred pressure data indicates a relatively high pressure behind the detonation wave, which gradually decreases as a result of expansion waves. Although all grid configurations except the one with no AMR exhibit similar pressure profiles, finer grids, especially those with 2 and 3 AMR levels, show relatively similar results. This suggests that for mesh resolutions featuring 2 or more AMR levels, the results become independent of the mesh. Therefore, the computational mesh featuring 2 levels of AMR ($\Delta_{min}$ = 0.0625 mm) is used in all numerical simulations carried out in this work. This aligns with the previous work of Chen et al. (2023) that used grid sizes of 0.1 mm.

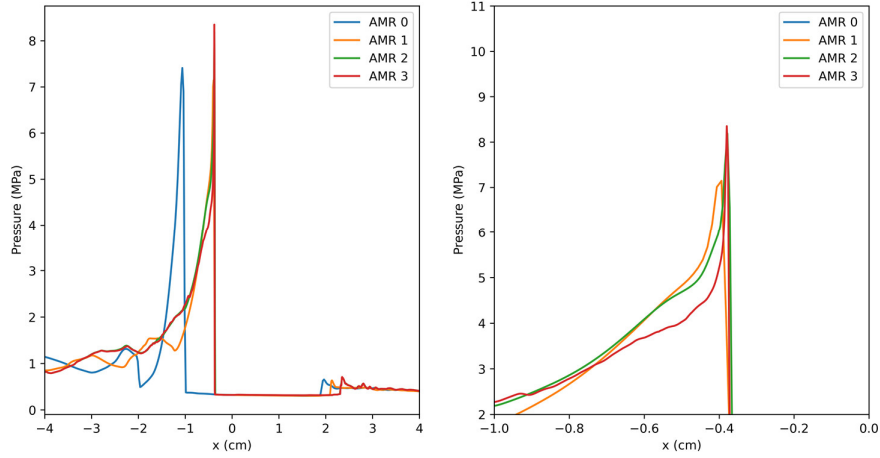

Figure 4. Pressure profiles along the x-direction and at y = 0.5 mm computed with four different meshes (AMR 0, AMR 1, AMR 2, and AMR 3) and a continuous inlet configuration. Right plot is a zoom of the left one.

## 3.2 RDE flow field structure

In this work, 5 different flow configurations of RDEs, identified as Case 1 to Case 5, were analyzed. In all these flow configurations, the total inlet pressure $P_0$ and the total temperature $T_0$ were fixed at 0.5 MPa and 300 K, respectively. Table 1 lists the number of nozzles that correspond to each Case, increasing from 15 inlets in Case 1 to 240 inlets in Case 5, such that each flow configuration has twice the number of inlets as the previous one.

Table 1  RDE flow configurations.

| Case | N° inlet nozzles | Injection width (mm) | Wall width (mm) | Self-sustained detonation |
|------|------------------|----------------------|-----------------|---------------------------|
| **1** | 15 | 6.000 | 4.000 | Failure |
| **2** | 30 | 3.000 | 2.000 | Yes |
| **3** | 60 | 1.500 | 1.000 | Yes |
| **4** | 120 | 0.750 | 0.500 | Yes |
| **5** | 240 | 0.375 | 0.250 | Yes |

Accordingly, this section compares the effect of the number of nozzles on temperature and pressure fields. The analysis focuses on time instants when RDWs become stable, except for the flow configuration featuring 15 inlets (Case 1), which experienced a detonation propagation failure. Notice that in this work a detonation wave is said to be stable when it propagates continuously and self-sustains over several wave cycles without significant decay of key parameters such as detonation velocity and mass flow rate. This definition aligns with others employed in previous research works, such as the work by Ullman and Raman (2024), where stability was observed after approximately three wave cycles, and by Chen et al. (2023), where stability was reached after about 0.35 ms, corresponding to roughly five wave cycles under similar geometric and initial conditions. More specifically, to ensure that detonation waves reach a stable state, the numerical simulations carried out in this work involved 20 wave cycles, and only the final five ones were considered for statistical analysis. Thus, the temperature and pressure fields shown for Case 1 correspond to an instant before the failure. For the remaining flow configurations (Case 2 to Case 5), the instantaneous temperature and pressure fields shown in Figure 5 and Figure 6, respectively, correspond to a time when both the RDEs have completed 15 cycles and the RDWs reach the center of the computational domain.

More specifically, Figure 5 illustrates the temperature distributions for all RDE flow configurations accounted for in this work (Table 1). As seen in this figure, in each flow configuration, after an initial transient period, a single detonation wave propagates stably from left to right. Besides, the typical two-dimensional RDE flow structures, which are characterized by a detonation wave, an oblique shock wave, a slip line, a contact surface, and a reactant filling region, are observed in all cases. Particularly, the shape of the refill layer observed in the numerical results obtained in this work is influenced by the striated nature of the reactants and the relatively low inlet pressure employed here. Similar outcomes were reported by Zhao et al. (2020), who assessed a continuous RDE configuration under varying inlet pressures (10 to 40 atm), revealing that the use of relatively low inlet pressures, such as 10 atm, result in highly distorted and wrinkled contact surfaces. The use in this work of an even lower inlet pressure of 0.5 MPa (~5 atm) also leads to non-uniform contact surfaces, driven by the interaction of these surfaces with reflected shock waves, which contributes to local instabilities in mass flow rates. From Figure 5 it is observed that the main differences in the temperature results obtained for each flow configuration lie indeed in the contact surface and the filling region, both influenced by the non-uniformity of the reactant region ahead of the detonation wave.

This last region comprises fresh unburned injected mixtures and trapped burned gas located within the intervals of the inlet nozzles. The discrete inlet nozzle configuration significantly influences the formation of this burned gas area, which is more prominent in cases with a relatively small number of inlets, Case 1 for instance, where unburned gases are completely segregated from the burned ones. In flow configurations featuring a relatively high number of inlet nozzles in turn, such as Case 4 and Case 5, this region of burned gases is minimized, and the RDE flow fields are similar to the one characterizing a continuous inlet (Figure 3), where the RDW propagates in a uniform area of injected gas mixture.

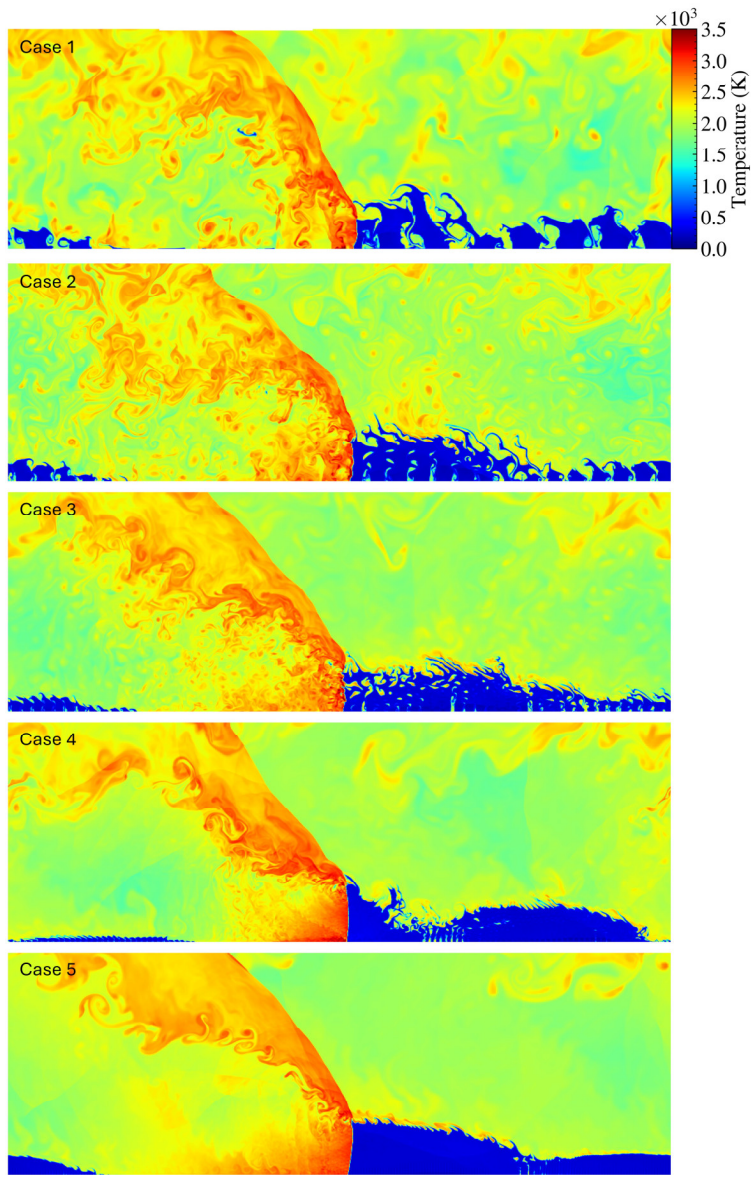

Figure 5. Temperature fields for Cases 1 to 5 (top to bottom).

As illustrated in Figure 5 and Figure 6, which shows pressure fields along with H2 mass fraction contours (black lines), the use of a relatively low number of inlet nozzles results in both more dispersed distributions of reactants and less stable detonation waves. Consequently, the RDE flow configurations corresponding to Cases 1, 2, and 3 exhibit highly banded distributions of reactants, where higher areas of burned combustion products are visible between the fresh premixed fuel injected into the RDE (Figure 5). The combustion products between the banded reactants cannot provide sufficient energy for the propagation of the detonation waves, resulting in weaker pressure detonation regions in the wave front (first three cases in Figure 6), leading to an unstable detonation front. The flow inhomogeneities generated by the discrete inlet nozzles also influence the intrinsic stability of the cellular detonation, leading to unstable frontal structures of the rotating detonation waves that differ from the one characterizing Case 5, which features a more uniform detonation wave front. These unstable detonation wave fronts (Figure 7) significantly decrease the peak pressure at the wave front, which ultimately affects the performance of the RDE. They also generate unburned fuel pockets behind the detonation wave, which will deflagrate rather than contribute to the detonation process, reducing thereby the efficiency of the RDE.

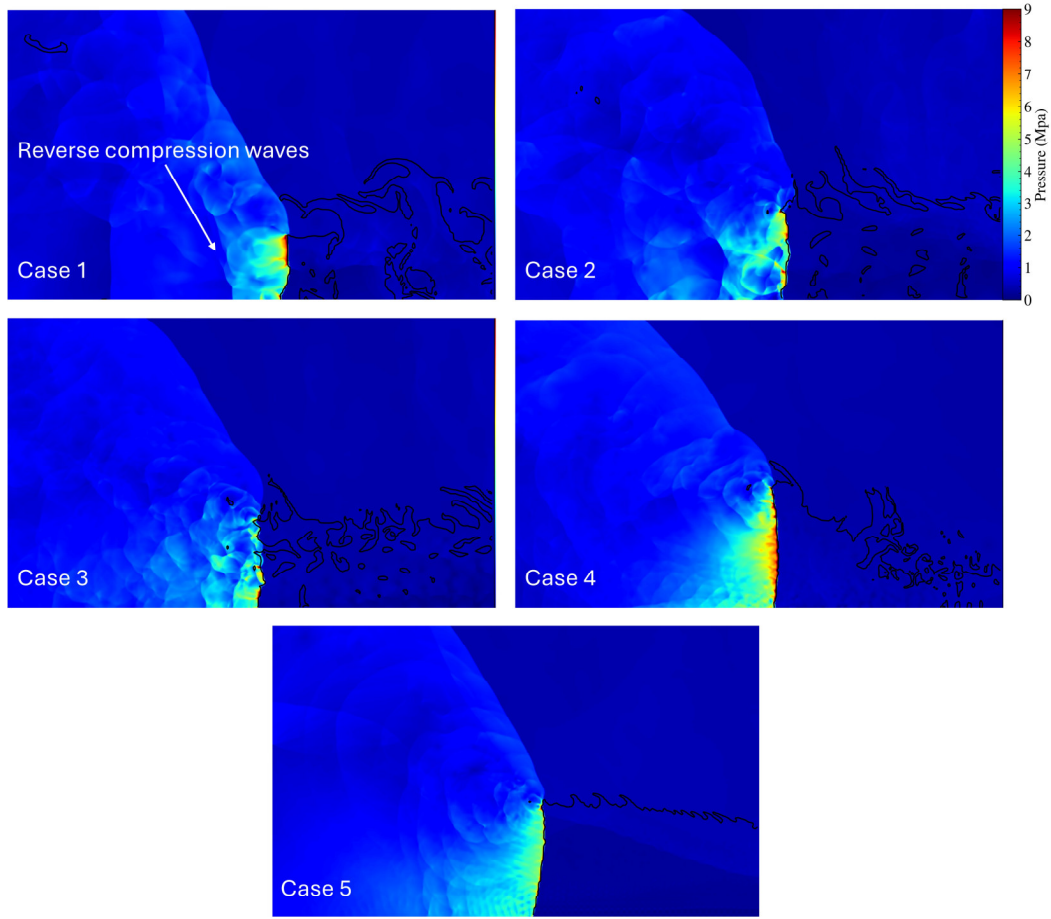

Figure 6. Pressure fields and H2 mass fraction contours (black lines) for all RDE flow configurations accounted for (Cases 1 to 5).

### 3.3 RDW stability characteristics

The failure in the detonation propagation that occurred in the RDE flow configuration featuring 15 inlet nozzles (Case 1) is attributed to the significant inhomogeneities present in the filling region characterizing this flow configuration. This aspect can be observed in Figure 5, which shows that, compared to the flow configurations featuring a larger number of inlet nozzles, in this case the increased regions of burned gases between the freshly injected premixed fuel contribute to instabilities, which manifest themselves as fluctuations in wave speed or peak pressure. This finding is further corroborated by the results shown in Figure 7, which shows the instantaneous pressure fields along with H2 mass fraction contours (white lines) for the flow configuration under discussion (Case 1) between 1.33 and 1.43 ms. Figure 7 highlights the fact that the decrease in pressure over time due to the referred flow inhomogeneities in the filling region results in a more distorted and weak detonation front, eventually leading to the extinction of the RDW. That is, despite some random local explosions within the flow, the detonation front does not recover and ends by disappearing. The results obtained in this work align with others discussed in previous studies (Liu et al., 2020; Wang et al., 2020), which found that reducing the inlet area ratio leads to a similar inhomogeneous region in front of the detonation wave, as observed in this work in cases with a relatively low number of inlet nozzles. It is worth noticing here that the mixing effect of the burned gases with the fresh mixture interrupts the layer of fresh mixture layer and causes the detonation wave to become unsustainable, eventually quenching it.

Figure 7 illustrates in particular the pressure drops occurring in the RDE flow configuration featuring 15 inlet nozzles and its decoupling from the RDW, which eventually lead to the RDW extinction. Indeed, with a higher flow uniformity in the region of fresh fuel mixture and burned products, the pressure wave is effectively reduced. Thus, at the first time instant (1.339 ms) shown in Figure 7, significant unburned H2 pockets begin to appear behind the detonation wave, a phenomenon not observed for instance in the flow configurations featuring 120 and 240 inlet nozzles (Cases 4 and 5). In addition, from 1.342 ms onwards, the pressure wave becomes weaker, and from 1.356 ms onwards, more pockets of unburned fuel appear behind the wave, due to the reduced capacity of the detonation wave to burn the mixture. The RDW stability is therefore closely related to the number of inlet nozzles employed. To achieve a more stable RDW, a greater homogeneity of the mixing zone is needed. For the remainder of this paper, only the RDE flow configurations presenting a self-sustained detonation (Cases 2 to 5) will be analyzed.

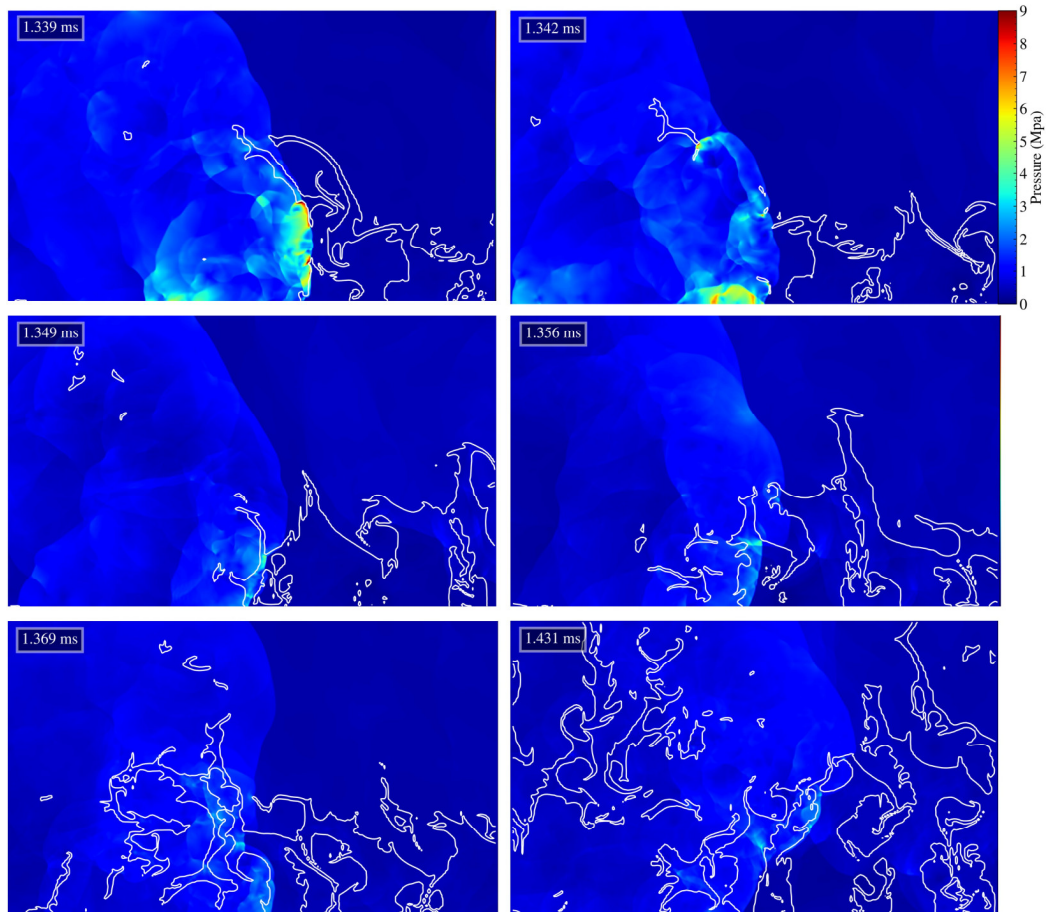

Figure 7. Pressure fields along with H2 mass fraction contours (white lines) for Case 1 (15 inlet nozzles) at different time instants.

It is worth noticing here that Lu and Braun (2014) observed that the interface where fresh reactants meet previously burned gases might destabilize the detonation wave by reducing its height or causing it to degenerate into deflagration. Additionally, Li et al. (2018) identified three key factors that cause instabilities at this boundary, (i) the Kelvin-Helmholtz (K-H) effect, (ii) the Rayleigh-Taylor (R-T) phenomenon, and (iii) the baroclinic torque. Due to the appearance of Kelvin–Helmholtz instabilities at the interface of the injected fuel, unburned gas pockets form and enter the junction between the detonation and oblique shock waves leading to strong explosions (Hishida et al., 2009). Accordingly, in this work, the RDE cases featuring a reduced number of inlets are more prone to the presence of these unburned gas pockets, suggesting a stronger influence of K-H instabilities in such cases. R–T instabilities occur in turn along an interface between two fluids featuring different densities and, specifically, when the lighter fluid is accelerated towards the heavier one. In the RDE cases with fewer inlets studied here, there is a greater density-related disparity between the fluids (as shown in Figure 5), leading to higher R–T instabilities. R–T instabilities can be understood as a result of the baroclinic torque, $\nabla \rho \times \nabla p$, represented by the vector product of density and pressure gradients, created by the misalignment of these gradients on the deflagration surface due to complex shock waves. Notice as well that in RDE cases with fewer inlets, the increased number of regions with density gradients amplifies the baroclinic torque. This amplification occurs because the spatial variation of density and pressure within these regions is heightened. Thus, the interaction between unburned gas pockets and detonation and oblique shock waves highlights the intricate dynamics and challenges of stable propagation of detonation waves within RDEs.

### 3.4 Effect of reverse compression shock waves

In all RDE flow configurations featuring 30 or more inlet nozzles, in addition to the rotating detonation wave, weak reversed compression shock waves are also observed. These reversed compression shock waves are shown in Figure 8, which describes modified pressure gradient fields (Liu et al., 2020; Meng et al., 2022). The referred compression shock waves arise because of the discontinuities in the fresh mixture and the region of burned products ahead of the detonation waves. More specifically, when detonation waves pass through combustion products, a minimal chemical reaction occurs. However, when detonation waves contact fresh reactants, a sudden chemical reaction occurs. These chemical reaction-related oscillations coming from the interaction between fresh and burned gases result in the referred reversed compression shock waves.

To distinguish weak compression waves from the detonation wave in the flow field, previous studies (Liu et al., 2020; Meng et al., 2022) have employed a modified pressure gradient parameter $\|\nabla p\| = \beta \exp\left(\frac{-k|\nabla p|}{|\nabla p|_{max}}\right)$, where $\beta$ and $k$ are two tunable coefficients. To obtain the results shown in Figure 8, a similar approach has been used in this work, accounting for coefficient values of $\beta = 1.0$ and $k = 100$. Figure 8 highlights in particular that the number of reverse compression shock waves increases with the decrease in the number of inlet nozzles. These compression shock waves interact with the contact surface in the mixing zone, leading to increased detonation front distortion, which could further destabilize the detonation front, as discussed in Section 3.3.

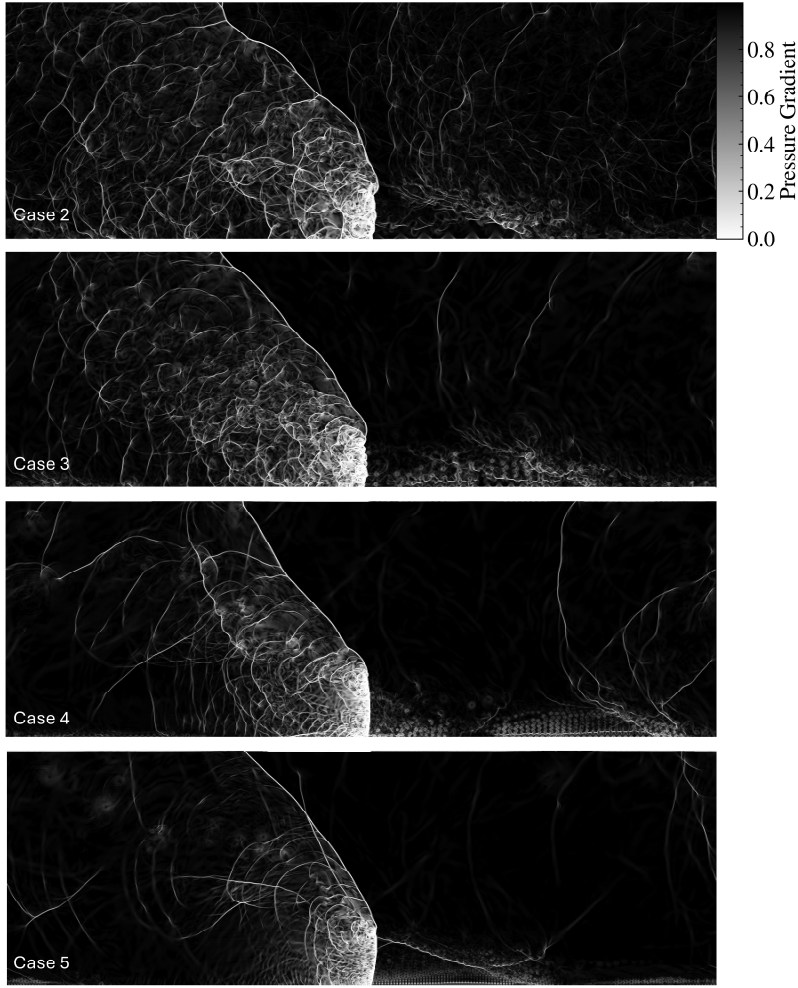

Figure 8. Pressure gradient $\|\nabla p\|$ fields for RDE flow configurations featuring 30 or more inlet nozzles (Cases 2 to 5).

Additionally, the referred reverse compression shock waves also influence the inflow boundary conditions, potentially causing instabilities in the local mass flow rates at the RDE inlet nozzles. Notice that in this work the inlet boundary condition is determined by both pressure $P_w$ and temperature $T_w$ inside the RDE, specifically in the nearest cell to the bottom boundary, in conjunction with the $P_0$ and $T_0$ inlet conditions. Indeed, when the $P_w$ values are higher than $P_0$, the reactants cannot be injected into the RDE, and the inlet behaves as a wall. In this context, a previous work (Zhao et al., 2020) has shown that for a total pressure ($P_0$) of 20 atm, the reverse compression waves do not significantly affect the fuel mass flow rates. However, for cases where the pressure $P_0$ is equal to 10 atm, these reversed shock waves not only influence the wrinkling of the contact surface, but also significantly alter the total mass rate entering the computational domain. In the RDE flow configurations studied here, where the pressure is as low as 5 atm, the impact of the reverse shock waves on the total mass flow rate is expected to be even more pronounced. It is worth noticing here that, because of the inlet pressure condition accounted for, the numerical results obtained by Zhao et al. (2020) for a 5 atm inlet pressure case cannot be directly compared to the ones obtained in the present work. This happens because Zhao et al. (2020) employed a sinusoidally varying inlet pressure, which led to the formation of multiple detonation waves. This flow configuration differs then from the one used here, where a constant inlet pressure of 5 atm was employed.

Consequently, Figure 9 shows the temporal evolution of the mass flow rate for the RDE flow configurations featuring 30 or more inlet nozzles (Cases 2 to 5). More specifically, Figure 9 shows the temporal and statistical analysis of the mass flow rate obtained from,

$$\dot{m} = \int_l \rho v dl \tag{8}$$

where $\rho$ refers to the density of the gas mixture, $v$ to the y-component of velocity, and $l$ is the cell size. Notice that in this figure, as well as in Figure 10 and Figure 11, the time axis has been normalized to wave cycles, using a mean wave speed of 1983 m/s. This wave speed was calculated as the average from Cases 2 to 5. In addition, the statistical data presented in Figure 9 and Figure 11 was derived from the last five wave cycles of the numerical simulations carried out.

This calculation was performed for all RDE flow configurations accounted for here, revealing that Case 2, which has the highest number of reverse compression shock waves, also features the lowest mass flow rate. It is worth mentioning that the temporal evolution of the mass flow rate associated with this flow configuration showed a relatively high standard deviation, with mass flow rates ranging from about 38 kg/ms to 64 kg/ms. This high variability in mass flow rate is mainly attributed to the increased number of the reverse compression waves.

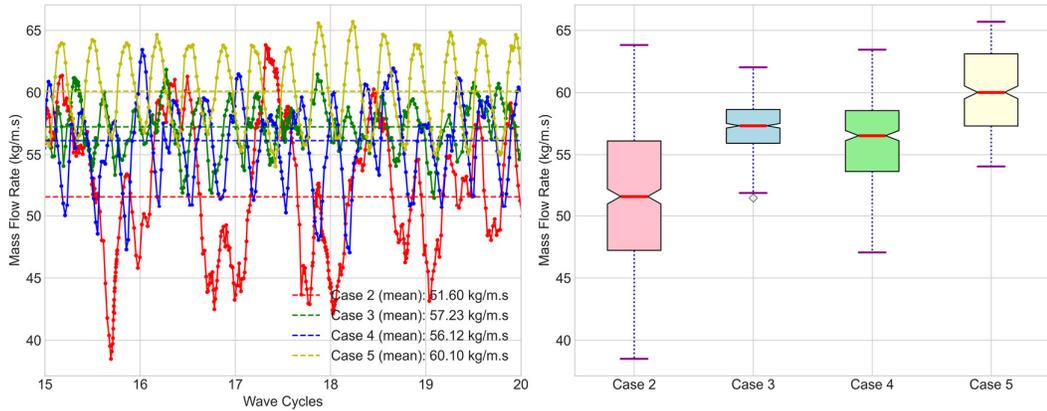

Figure 9. Temporal and statistical analysis of total mass flow rate for RDE flow configurations featuring 30 inlet nozzles or more (Cases 2 to 5).

As observed in Figure 9, as the number of inlet nozzles increases, so does the mass flow rate, accompanied by a decrease in the standard deviation. This trend is mainly attributed to the decrease in the number of reverse shock compression waves, resulting in more stable RDWs. However, there is a small discrepancy in the results that characterize Case 3, which has a mean mass flow rate of 57.23 kg/ms, slightly higher than the 56.12 kg/ms associated with Case 4. Furthermore, the standard deviation of Case 3 is lower than that of Case 4. Despite these small discrepancies, the overall trend indicates an increase in mass flow rate with the number of inlet nozzles, as exemplified by Case 5, which shows the highest and most consistent mass flow rate values.

It is worth noticing here that there are previous works where similar RDE mass flow rate related results are discussed. For instance, in the work by Ullman et al. (2024), which examines the effect of varying the number of inlets on RDE performance accounting for an inlet pressure of 1 MPa, it is observed that the mass flow rate decreases with the increase in the number of injectors, which differs from the findings of the present study. This discrepancy is attributed to the fact that in the referred study multiple detonation waves are formed, which as emphasized by Eq. (5) obstruct the injector inlets and reduce the effective mass flow rate. In contrast, in the present study, across all flow configurations accounted for, only a single detonation wave is consistently observed. This result comes primarily from the relatively low inlet pressure used in the numerical simulations carried out here. Notice as well that, for the same flow configuration and an inlet pressure of 0.5 MPa, Chen et al. (2023) also reported the formation of a single detonation wave, regardless of the variation of the fuel equivalence ratio. They noticed indeed that multiple detonation waves emerge only when the inlet pressure is increased to 1.2 MPa. This implies that a sufficiently high fresh mixture supply, proportional to the inlet pressure, is necessary for the formation and propagation of multiple detonation waves, as the operating mode transition threshold is primarily determined by mass flux affecting the number of resulting detonation waves (Mundt et al., 2024). In the present work, the increase in the number of injectors leads to a more uniformly distributed flow across the RDE annulus, reducing local flow blockages and ensuring that each injector effectively contributes to the overall mass flow rate.

## 3.5 Detonation wave velocity

The velocity of the detonation waves traveling in the RDE flow configurations considered in this work is computed based on the physical location of the RDWs. Accordingly, to track RDW locations in the computational domain, the obtained numerical results are first filtered accounting for the heat release rate (HRR). Indeed, following previous works on RDE involving H2/air mixtures (Zhao et al., 2020; Zhao & Zhang, 2020), a threshold value of HRR equal to 10e13 J/m3/s is utilized, so larger values than this threshold determine the RDW locations. After carrying out the referred filtering process and determining the RDW locations, both height and position of the detonation waves are extracted and averaged over 5 cycles (5 flow-through time) of the RDWs. More specifically, the RDW height is determined by the y-direction distance from the inlet to the end where the HRR exceeds the threshold value. Similarly, the position of the detonation wave is determined as the x-direction distance from the left end boundary to the RDW location.

Figure 10 shows thus the temporal evolution of the height and position of the RDWs for RDE flow configurations featuring 30 or more inlet nozzles (Cases 2 to 5). Notice that, to obtain the results shown in this figure, the RDW locations have been adjusted in all cases to start from the same point in space, facilitating in this way the comparison processes. The referred figure shows that the RDW heights decrease from Case 2 (30 inlet nozzles) to Case 5 (240 inlet nozzles), with time-averaged values of 1.354, 1.204, 1.129, and 1.128 cm, respectively. It is worth noticing here that increasing the number of reactant jets increases the total surface area of reactants exposed to recirculating product gases, thereby facilitating increased deflagration and a reduction in reactant fill heights. With fewer inlet nozzles, the mixing between the burned products and the fresh premixed fuel is also significantly reduced, leading to higher reactant fill heights. This effect is evident in the RDE flow configurations featuring 120 and 240 inlet nozzles, Cases 4 and 5, respectively, which have a minimal region of combustion products between the incoming fresh fuel, so they exhibit similar RDW heights. It is also worth noticing that, in the flow configuration featuring 30 inlets (Case 2), the RDW shows relatively large amplitude oscillations, indicating its low stability. In contrast, the flow configuration featuring 240 inlets (Case 5) exhibits a more stable height.

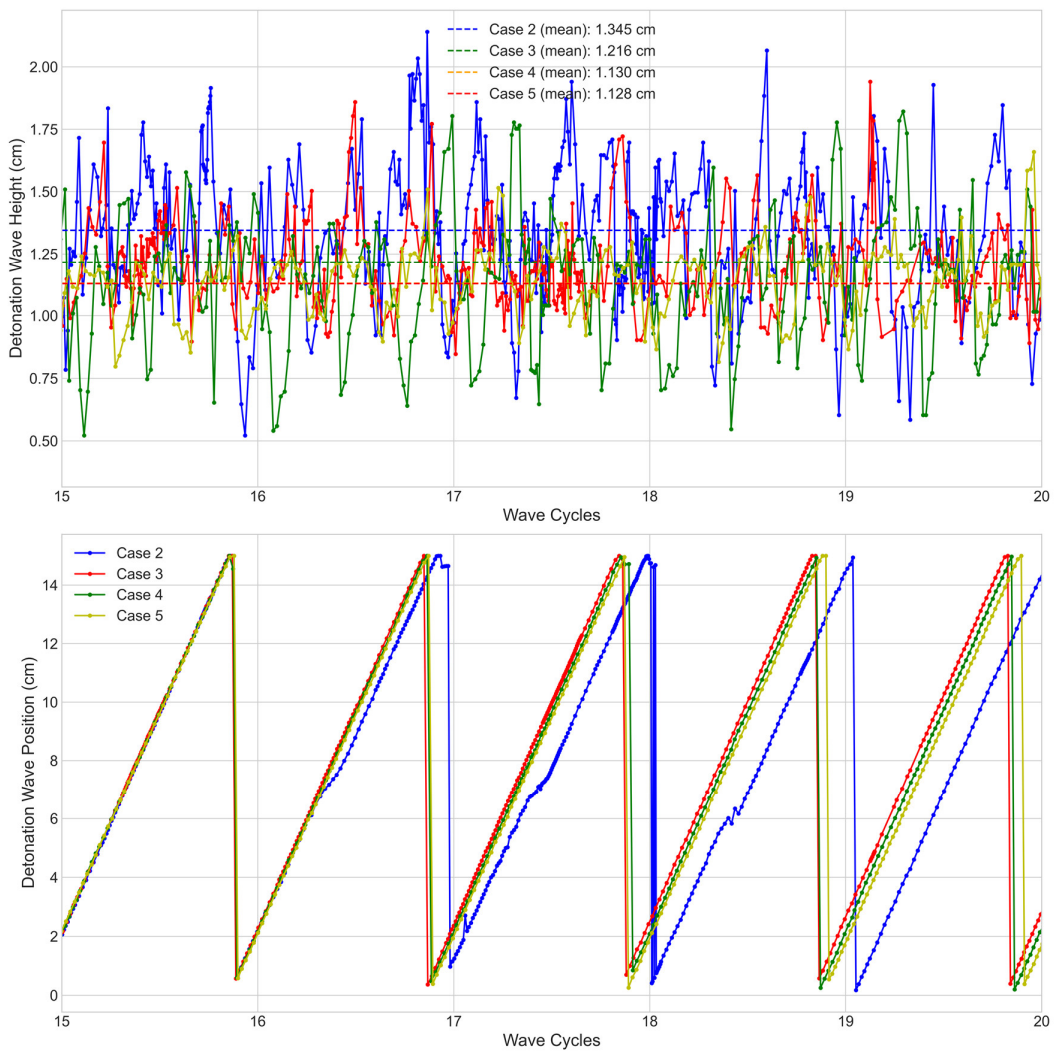

Figure 10. Temporal evolution of RDW height and position for RDE flow configurations featuring 30 inlet nozzles or more (Cases 2 to 5).

Regarding the variation over time of the RDW positions shown in Figure 10, it is worth noticing first that, contrarily to what happen in the other flow configurations studied here, Case 2 (30 inlet nozzles) displays some phase shifts between the 16 and 19 wave cycles. This phase shifts arises from flow instabilities similar to those ones observed in Case 1 (Section 3.3). However, unlike Case 1, the RDW in Case 2 did not extinguish but instead regained intensity and remained stable. From the four RDE flow configurations discussed in this section, Case 2 emerges as the most critical one due to its pronounced RDW instabilities, attributed to the lower number of inlet nozzles. For the remaining configurations (Cases 3 to 5), the differences in RDW positions are minimal, suggesting similar detonation wave velocities across all these cases. However, there are still some small differences in detonation wave speeds (Figure 11). For instance, the detonation wave speed in Case 5 is lower than in Cases 3 and 4. Notably, Case 3 demonstrates the highest wave speed, as it exhibits the most significant displacement over the five wave cycles.

The detonation velocity $V_D$ was computed in this work by determining the difference in terms of RDW position between two time instants and dividing it by the time variation itself. Furthermore, the velocity deficit $D_V$ (in %) relative to the ideal Chapman-Jouguet velocity $V_{CJ}$ was evaluated using $D_V = \frac{(V_{CJ} - V_D)}{V_{CJ}} * 100$. The ideal Chapman-Jouguet (CJ) velocity was calculated here using Cantera (Goodwin et al., 2021) via the Shock and Detonation Toolbox library (California Institute of Technology, 2024), based on $T_{cr}$, $P_{cr}$, and the same chemical kinetic mechanism employed in the numerical simulations carried out in this work. Accordingly, Figure 11 compares the detonation velocities characterizing the RDE flow configurations featuring 30 inlet nozzles or more (Cases 2 to 5), along with the velocity deficit compared to the CJ velocity of 2034.93 m/s. Overall, as shown in this figure, the detonation velocities in all flow configurations are similar and close to the theoretical CJ one, with velocity deficits below 3.2% in all cases.

These findings agree with previous ones obtained by Fujii et al. (2017) accounting for methane as a fuel and focusing on detonation velocity analyses. The referred study concluded indeed that the burned gases in front of the detonation waves do not significantly affect the propagation velocities of detonation waves, a finding consistent with the numerical results obtained here. It is worth noticing however that, since parasitic deflagration and recirculating product gases both (i) partially consume the reactants (thereby leaving less chemical energy available to support detonations) and (ii) heat the mixture ahead of detonations (which increases the local speed of sound and thus decreases the compression ratio that can be achieved across detonation waves), it is unlikely that the presence of product gases have no effect on detonation waves. Therefore, there must be other physical processes at play that mitigate the deleterious effects of product gases and support the speed of detonation waves. As noticed before, similar observations were made by Fujii et al. (2017), who varied the number of fuel injectors from 10 to 500 and obtained a consistent detonation wave speed close to the CJ one across all flow configurations analyzed. Therefore, this issue needs to be further investigated in future work. Notice that the box plot shown in Figure 11 (bottom) represents the dispersion of the detonation velocities in each of the RDE flow configurations discussed here. From this figure Cases 2, 3, and 4 have more dispersion compared to Case 5, indicating a greater consistency in detonation wave velocities, due to a more stable detonation front, with a higher number of inlet nozzles. With 240 inlets indeed, despite having the second highest velocity deficit (2.93%), Case 5 exhibits the smallest variation in detonation wave velocity.

In addition, as shown in Figure 11, due to the substantial amount of burned gas between the fresh fuel mixture observed in Case 2 (Figure 5), its detonation velocity is the lowest among all the flow configurations analyzed. The reduction in wave speed with fewer inlet nozzles can be attributed to several interrelated factors. For instance, with fewer inlet nozzles, the spacing between the reactant jets increases, leading to a less uniform distribution of reactants in the combustion chamber. This larger spacing results in less consistent interactions between detonation waves and reactant jets, contributing to increased wave instability. Moreover, a reduced number of inlet nozzles also leads to a less effective mixing of the injected reactants with the surrounding gases. This inefficient mixing creates in turn regions with varying reactivity within the chamber, causing fluctuations in wave propagation speed and contributing to wave instability. Additionally, the uneven distribution of reactants and burned gases leads to variations in reactivity within the chamber. Therefore, detonation waves encountering regions featuring different reactivity may experience inconsistent wave strengths and increased wave instability. Besides, with fewer nozzles, the detonation waves may undergo more significant interactions and reflections within the chamber, which may alter wave speed and stability, leading to the observed differences in the numerical predictions carried out. Whereas for the other flow configurations studied in this work the detonation velocity does not change much, it becomes more uniform with the increase in the number of inlet nozzles.

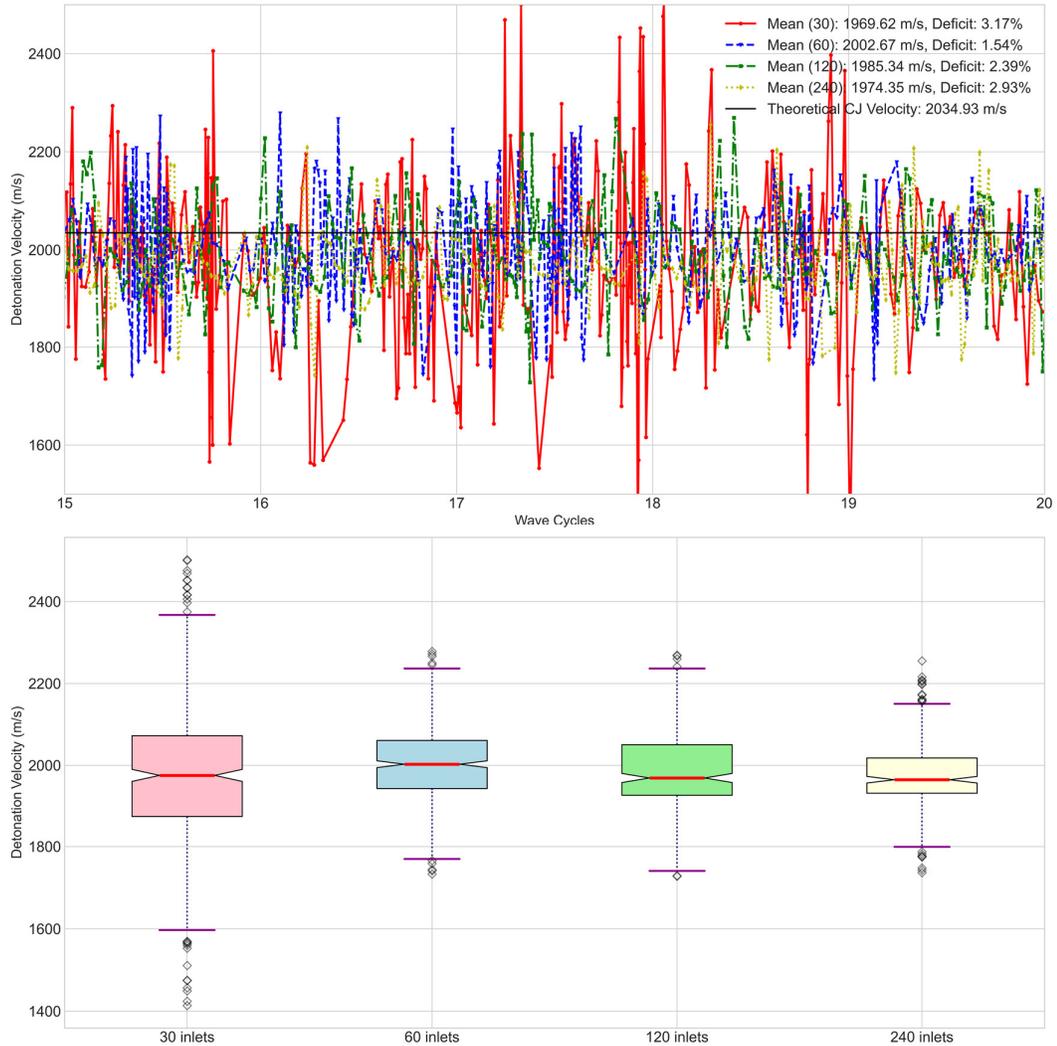

Figure 11. Temporal and statistical analysis of detonation wave velocity for RDE flow configurations featuring 30 inlet nozzles or more (Cases 2 to 5).

## 3.6 RDE performance

The performance of RDEs is typically assessed in terms of two key parameters, (i) the specific impulse and (ii) the specific thrust (Schwer & Kailasanath, 2011). The specific impulse $I_{sp}$, through the quantification of the thrust produced per unit weight of fuel consumed, serves as a measure of how effectively the engine uses the fuel to generate thrust. Conversely, by quantifying the thrust produced per unit weight of oxidizer consumed, the specific thrust $F_{sp}$ provides a measure of how efficiently the engine uses the oxidizer to generate thrust. These two parameters are computed in this work using the following expressions (Chen et al., 2023),

$$I_{sp} = \frac{F}{\dot{m}_{H_2} g} \tag{10}$$

$$F_{sp} = \frac{F}{\dot{m}_{air}} \tag{11}$$

where $\dot{m}_{H_2} = \int_l \rho_{H_2} v dl$ is the mass flow rate of H2, $\dot{m}_{air} = \int_l \rho_{air} v dl$ is the mass flow rate of air, $g$ represents gravity, and $F$ stands for thrust. As observed from the last two equations, to compute the specific impulse and the specific thrust, the thrust at the combustor outlet is required. This last parameter is determined in this work as follows (Chen et al., 2023),

$$F = \int_l (\rho v^2 + P - P_\infty) dl \tag{9}$$

where $P_\infty$ is the ambient pressure, equals to 0.1 MPa here, and $P$ is the pressure at the RDE outlet. To analyze it further, the thrust is usually divided into two components, one related to the momentum thrust $\int_l (\rho v^2) dl$, and the other to the pressure thrust $\int_l (P - P_\infty) dl$.

Accordingly, Figure 12 illustrates the mean values of the two thrust components (momentum thrust and pressure thrust) and the total thrust for the RDE flow configurations featuring 30 or more inlet nozzles (Cases 2 to 5), with the error bars indicating the standard deviation of the computed values. As noticed from this figure, the main thrust contributor in all flow configurations studied here is the pressure thrust, contributing to about two-thirds of the total thrust. For all cases however, this pressure thrust remains relatively constant, around 65 kN/m, except for Case 5, which shows the highest value of 66.76 kN/m. In contrast, the momentum thrust exhibits the largest variations, with mean values of 34.33, 40.39, 39.20, and 42.48 kN/m for the RDE flow configurations corresponding to Cases 2, 3, 4, and 5, respectively. While the pressure thrust (green bars) shows an increase with the number of injectors, indicating the presence of stronger detonation waves, its relative contribution to total thrust changes is smaller compared to that of the momentum thrust. This suggests that the impact of the mass flow rates on the momentum thrust is more significant, making it the dominant factor in the overall increase in total thrust. The referred large variations in momentum thrust are directly linked to the y-component of the flow velocity at the RDE's outlet (Figure 13). Therefore, being the sum of the momentum and pressure thrusts, the total thrust mirrors the trend of the momentum thrust, as the pressure one remains relatively constant. Case 5 (240 inlet nozzles) features the highest total thrust, about 108.70 kN/m, primarily due to its highest momentum thrust. Conversely, Case 2 (30 inlet nozzles) shows the lowest total thrust, 99.63 kN/m, attributed to its lowest momentum thrust. These results underscore the significance of both momentum and pressure thrust components in determining the overall thrust performance of rotating detonation engines, with momentum thrust playing a dominant role.

In general, an increase in the number of inlet nozzles leads to a higher total thrust. This is because a higher number of inlets results in an increased mass flow rate entering the RDE (Figure 9), indicating that thrust primarily depends on the amount of mixture entering the RDE. The slight differences between Cases 3 and 4 can be attributed to Case 3 having a mass flow rate (57.23 kg/s) slightly higher than that characterizing Case 4 (56.12 kg/s), resulting in a higher total thrust for Case 3 (104.53 kN/m vs. 103.92 kN/m). In terms of standard deviation, the RDE flow configurations associated with Cases 2 to 5 feature values of 8.69, 3.63, 6.63, and 5.28, respectively. As expected, Case 2 shows the most unstable thrust. Surprisingly, Case 3 exhibits a standard deviation lower than Case 5, which featured a more stable detonation front. This result suggests that, despite Case 3 having a relatively high standard deviation in the mass flow rate, it still manages to maintain a relatively stable pressure and velocity at the RDE outlet.

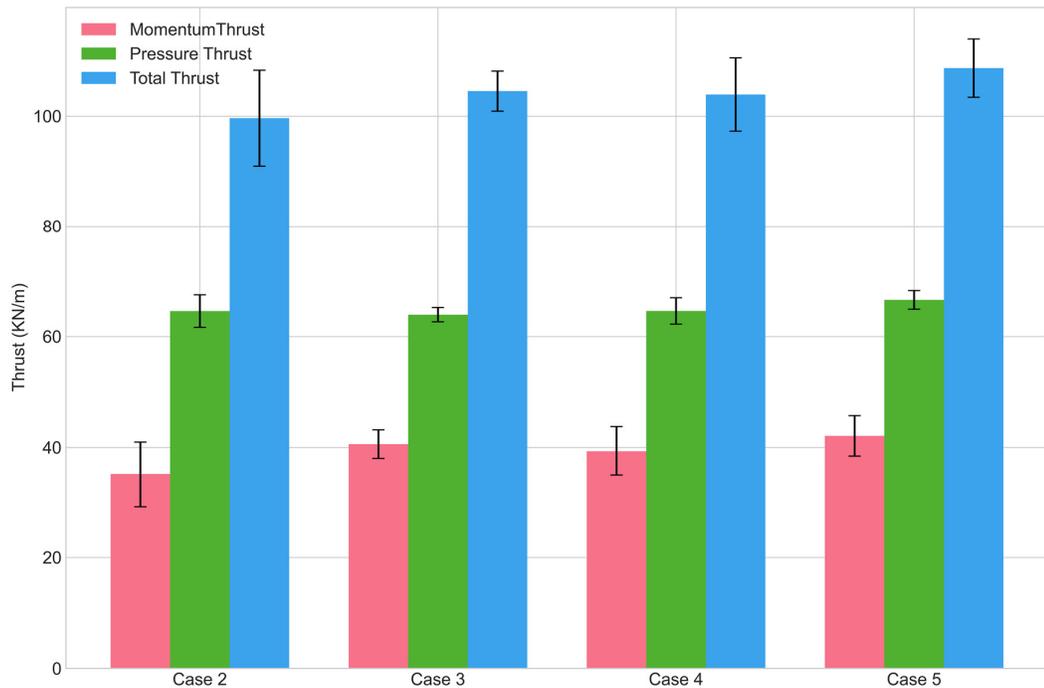

Figure 12. Comparison of mean thrust components and total thrust for RDE flow configurations featuring 30 inlet nozzles or more (Cases 2 to 5).

By describing the variations of the mean pressure, the y-component of the flow velocity, and the H2 mass fraction along the detonation height direction, Figure 13 allows carrying out further analyses of the RDE thrust results. Notice that, following previous studies (Liu et al., 2020; Zhao et al., 2021), the results shown in this figure were obtained through volume averaging (for the pressure) and mass averaging (for H2 mass fraction and y-velocity component) along the x-direction, and time averaging over the last five wave cycles of the numerical simulations performed. It is worth noticing as well that, as indicated by the horizontal dashed lines included in Figure 13, the mean heights of the RDWs characterizing Cases 2, 3, 4, and 5 are about 1.345, 1.216, 1.130, and 1.128 cm, respectively. From Figure 13 is first observed that the mean H2 mass fractions are initially high near the inlet nozzles but they rapidly decrease downstream until they reach the mean detonation height. It can be also noticed from this figure that, at 0 cm of height, Case 2 featuring an unstable detonation front exhibits the lowest value of the H2 mass fraction. The primary reason for this is the substantial presence of combustion products interspersed in the fresh fuel mixture. Although Case 2 has a reduced mean H2 mass fraction, it exhibits the most gradual hydrogen consumption rate. This can be attributed to the instabilities in the detonation front that limit the consumption of the hydrogen. Conversely, as it contains a minimal area of burned products between the incoming fresh fuel, Case 5 starts with the highest H2 mass fraction, behaving thus like a RDE with a continuous inlet configuration. However, it reaches the same mass fraction value as Cases 4 and 3 when it attains its detonation height. Case 4 behaves almost identically to Case 3, exhibiting similar H2 mass fraction profiles. Despite Case 4 having a higher initial H2 mass fraction, due to a more uniform filling region, it seems that Case 4 burns the fuel more uniformly across the height of the RDW. In addition, Case 3 consumes more H2 at lower heights, but in the end, both cases (3 and 4) reach the same H2 mass fractions when they attain their detonation wave heights.

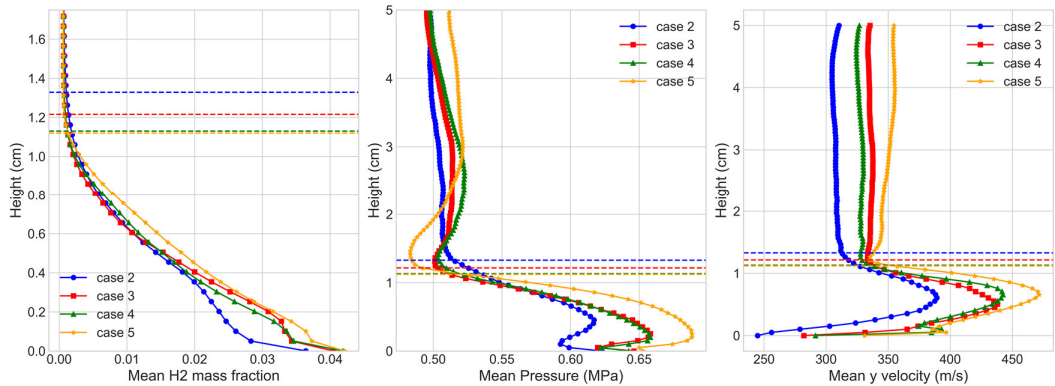

Figure 13. Mean profiles of H2 mass fraction, pressure, and y-velocity as a function of chamber height for RDE flow configurations featuring 30 inlet nozzles or more (Cases 2 to 5).

In terms of mean pressure, compared to the other RDE flow configurations studied here, Case 5 initially exhibits a significant increase in pressure at heights lower than its RDW height. This is primarily attributed to the increased number of inlet nozzles, which intensify the detonation wave, leading to higher pressure values. Notice however that this pressure subsequently decreases to its minimum at a height slightly above the RDW height. After reaching the minimum pressure value, Case 5 pressure increases again achieving the highest pressure value at the RDE exit. This behavior results in Case 5 having the highest pressure thrust. In turn, Case 2 has the lowest initial pressure value, but eventually achieves the same exit pressure as Cases 3 and 4. Indeed, Cases 3 and 4 exhibit similar pressure trends, with the main distinction being that Case 4 experiences a slightly higher pressure increase than Case 3 between the heights of 2 and 3 cm. This goes hand in hand with the fact that within the same height range, the velocity of Case 4 is lower compared to Case 3. Therefore, for Case 4, it can be said that there is a compensation of the pressure increase with a decrease in velocity. In the end, Cases 4 and 3 reach the same pressure at the exit of the RDE. In terms of exit velocity however, they do not obtain similar values. This is where Case 3 outperforms Case 4 in terms of thrust, since Case 3, by increasing its y-velocity, exits with a higher momentum thrust. Also, the performance enhancement of Case 3 relative to Case 4 is largely contingent on the greater amount of fuel introduced into Case 3 compared to Case 4.

Finally, Figure 14 shows the specific thrust and specific impulse obtained for all RDE flow configurations featuring 30 or more inlet nozzles (Cases 2 to 5). These results reveal a consistent reduction in both specific impulse and specific thrust as the number of inlet nozzles increases. Although maximizing the number of inlets results in maximum total thrust, these RDE flow configurations also feature higher fuel and air consumption rates. Despite the increase in the total mass flow rate, from 52.29 kg/s.m to 60.07 kg/s.m (14.87% increase) when transitioning from 30 inlets to 240 inlets, the improvement in specific performance is relatively small, as thrust values increase from 98.83 kN/m to 109.24 kN/m (10.5% increment). Except for Case 3 (lowest standard deviations for both specific impulse and specific thrust), a noticeable improvement associated with increasing the number of inlet nozzles is the achievement of more uniform thrusts attributed to more stable RDWs. This is emphasized in the results shown in Figure 14 as the standard deviation in Case 2 is approximately double that of Case 5. Case 3 featuring 60 inlets only exhibits however a lower standard deviation. Despite having a non-uniform total mass flow rate, it seems that this RDE flow configuration managed to achieve a more uniform thrust.

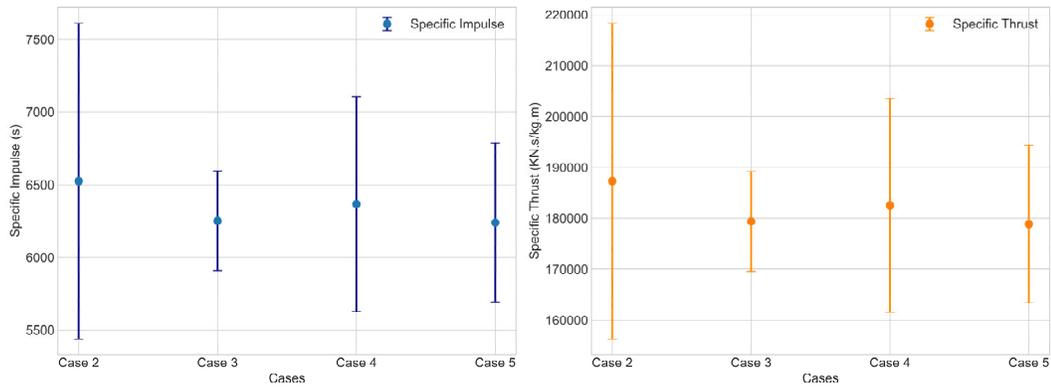

Figure 14. Mean specific impulse (left) and mean specific thrust (right), including error bars, for RDE flow configurations featuring 30 inlet nozzles or more (Cases 2 to 5).

## 4   Conclusions

In this work, rotating detonation engines (RDEs) fueled by premixed stoichiometric hydrogen/air mixtures were explored in depth through two-dimensional numerical simulations including a detailed chemical kinetic mechanism. By varying the number of discrete inlet nozzles, the influence of reactant non-uniformities on the dynamics of the associated detonation processes was particularly analyzed. The numerical results presented here provide several insights about the fundamental combustion science processes that govern the performance of RDEs when varying the number of inlet nozzles. For instance, they reveal that the flow field structure and the stability of rotating detonation waves (RDWs) are significantly affected by the number of inlet nozzles. Specifically, a higher number of inlets lead to a more uniform filling region, which results in more stable RDWs. However, the total thrust generated is primarily dependent on the amount of fuel mixture entering the RDE, rather than the number of inlets.

In addition, RDE flow configurations featuring a relatively low number of inlets, such as the 15-inlets one, exhibit higher RDW instabilities. The referred instabilities are characterized by non-uniform reactant distributions, which often lead to the formation of weaker pressure regions and pockets of unburned fuel. These factors adversely impact the stability of RDWs, leading eventually to their quenching. The results obtained here also unveils the presence of reverse compression waves that contribute to flow field distortions, particularly in RDE configurations including a relatively small number of inlet nozzles. Notice that such flow distortions can destabilize the RDWs and result in fluctuations in mass flow rates. The obtained results also show that the detonation velocities characterizing all RDE flow configurations studied here are similar and close to the theoretical Chapman-Jouguet one, with minor deficits ($< 3.2\%$).

Finally, the numerical results obtained in this work also highlight the importance of both momentum and pressure thrust components in determining the RDE overall thrust. While the pressure thrust remains relatively constant across all RDEs, the momentum thrust shows significant variations, thereby influencing the total thrust. In particular, higher numbers of inlets result in increased total thrusts, primarily driven by the increased momentum thrusts. However, RDE specific performance metrics such as the specific impulse and the specific thrust does not change significantly with the increase in the number of inlet nozzles because the increase in thrust also involves increased mass flow rates. Summarizing, achieving a balance between increasing the mass flow rate and achieving a uniform thrust for improved overall performance of the system requires careful consideration of the number of inlets. Although a higher number of inlet nozzles enhances the total thrust and the RDW stability, optimizing specific performance metrics requires further investigation into both fuel-air mixing and detonation processes within RDEs.

## 5   Acknowledgments

This work has been supported by the US Army Research Laboratory and the US Air Force Office of Scientific Research (AFOSR) under Research Grant No. W911NF-22-1-0275. Luis Bravo was supported by the US Army Research Laboratory 6.1 Basic research program in propulsion sciences. The computations were supported in part by CPUs at Purdue Anvil through allocation MCH240024 from the Advanced Cyberinfrastructure Coordination Ecosystem: Services & Support (ACCESS) program, which is supported by National Science Foundation grants #2138259, #2138286, #2138307, #2137603, and #2138296.